\documentstyle[aps,preprint,epsf,amssymb]{revtex}

\newcommand{\di}{\mbox{$i\!\!\not\!\!D$}}
\newcommand{\nub}{\mbox{$\overline{\nu}$}}

\newcommand{\ssi}{\mbox{$\langle\overline{\psi}\psi\rangle$}}
\newcommand{\nni}{\mbox{$n_{\overline{I}}$}}

\begin{document}
\preprint{OUTP-98-74P}
\title{On the spectral density from instantons in quenched QCD}
\author{U. Sharan \and M. Teper\thanks{ujji@thphys.ox.ac.uk, teper@thphys.ox.ac.uk}
({\it UKQCD Collaboration})}
\address{Theoretical Physics, University of Oxford,
1 Keble Road, Oxford, OX1 3NP, U.K.}
\maketitle

\begin{abstract}
We investigate the contribution of instantons to the eigenvalue
spectrum of the Dirac operator in quenched QCD.  The instanton
configurations that we use have been derived, elsewhere, from cooled
SU(3) lattice gauge fields and, for comparison, we also analyse a
random `gas' of instantons.  Using a set of simplifying
approximations, we find a non-zero chiral condensate. However we also
find that the spectral density diverges for small eigenvalues, so that
the chiral condensate, at zero quark mass, diverges in quenched
QCD. The degree of divergence decreases with the instanton density, so
that it is negligible for the smallest number of cooling sweeps but
becomes substantial for larger number of cools. We show that the
spectral density scales, that finite volume corrections are small and
we see evidence for the screening of topological charges. However we
also find that the spectral density and chiral condensate vary rapidly
with the number of cooling sweeps -- unlike, for example, the
topological susceptibility. Whether the problem lies with the cooling
or with the identification of the topological charges is an open
question. This problem needs to be resolved before one can determine
how important is the divergence we have found for quenched QCD.
\end{abstract}

\section{INTRODUCTION}

There have been a number of recent lattice calculations that attempt
to determine the instanton content of the vacuum in SU(2) \cite{Qsu2}
and SU(3) \cite{Qsu3,DSMT} gauge theories. One motivation has been to
make contact with phenomenological instanton models
\cite{Shuryak,Diak}.  Another aim has been to test certain long-held
theoretical ideas, such as that instantons might play an important
role in chiral symmetry breaking \cite{Shuryak,Diak,Qssb}.

To identify the instanton content of fully fluctuating vacuum gauge
fields (the latter obtained from lattice Monte Carlo simulations) is
far from trivial.  Indeed it is not entirely clear to what extent it
is either meaningful or possible. Current techniques involve
smoothening the lattice gauge fields on short distances and then using
some pattern recognition algorithm to resolve the topological charge
density into an ensemble of overlapping (anti)instantons of various
sizes and positions. At present there is only some agreement between
the results of the different approximate methods being used
\cite{Qsu3,DSMT}.  Thus all these calculations should be regarded as
exploratory.

In this paper we shall focus on the calculations in \cite{DSMT}.
There the smoothing of the rough gauge fields was achieved by a
process called `cooling' \cite{cool}.  This is an iterative procedure
just like the Monte Carlo itself except that the fields are locally
deformed towards the minimum of the action (or some variation
thereof). Although some quantities, such as the topological
susceptibility, are insensitive to the amount of cooling (within
reason) this is not the case for the number, size distribution and
density of the topological charges. Whether this leads to a real
ambiguity for physical observables is an important question. It might
be that the ambiguity is only apparent and that fermionic observables
calculated in these cooled instanton background fields do not show
much variation with cooling. For example it might be the case that the
instantons that disappear with cooling are highly overlapping $Q=\pm
1$ pairs which contribute no small modes to the Dirac operator.  In
that case the spectrum of small modes would be insensitive to cooling,
and so would various fermionic observables such as the chiral
condensate. How these physically important small modes actually depend
on cooling is the main question we address in this paper.

One might think that the reasonable way to approach all these
questions would be to perform calculations directly on the cooled
lattice fields using lattice versions of the Dirac operator. Although
such explicit calculations do show that it is the instanton (would-be)
zero modes that drive chiral symmetry breaking \cite{SHMT}, one also
finds that lattice artefacts spoil the mixing of the instanton
near-zero modes \cite{SHMT} and this makes it difficult to draw
reliable conclusions for the continuum limit. (Although very recent
work with domain-wall fermions \cite{wallferm}, and with related
lattice fermions \cite{newferm} suggests a promising avenue for
progress.)

In this paper we shall calculate the interesting low lying eigenvalue
spectrum of the Dirac operator by constructing an explicit matrix
representation for a given background field of overlapping instantons
and anti-instantons. Given our ignorance of the full structure of the
vacuum, such a method is necessarily approximate. Our approach will be
to simplify the details as much as possible while still incorporating
both the important symmetries of the problem and (in the model
calculations) the long-distance clustering properties of an ensemble
of topological charges.  Our expectation is that this should ensure
that our answers are qualitatively reliable \cite{usprep}.  Some
preliminary results using this method for configurations of instantons
and anti-instantons distributed at random have been given elsewhere
\cite{ndmt,usmt}.

Let $\lambda_{n}[A]$ be an eigenvalue of the Dirac operator for some
given gauge field configuration $A: \di[A]\phi_{n}(x) =
\lambda_{n}[A]\phi_{n}(x)$ defined in some space-time volume $V$. We
know via the Atiyah-Singer Index theorem \cite{atsi} that the
eigenvalue spectrum will contain

\begin{equation}
\label{eq:atsi}
Q[A] = n_{-} - n_{+}
\end{equation}
exact zero modes, where $Q[A]$ is the winding number of the gauge
field configuration and $n_{\pm}$ are the number of zero eigenmodes
with positive/negative chirality. In the chiral limit a zero
eigenvalue leads to a zero determinant; so gauge field configurations
with non-trivial winding number are suppressed by the light quarks of
full QCD. This suppression is lost in quenched QCD where the fermion
determinant is set to unity. Let $\rho(\lambda; A) =
\sum_{n}\delta(\lambda - \lambda_{n}[A])$ be the spectral density of
the Dirac operator. One can relate the chiral condensate to the
spectral density via

\begin{equation}
\label{eq:chico}
\ssi = \lim_{m \rightarrow 0}\
i\int_{0}^{\infty}\frac{2m\nub(\lambda,m)}{\lambda^{2} + m^{2}} ,
\end{equation}
where $\nub(\lambda, m) = \lim_{V \rightarrow \infty} \nu(\lambda, m)$
and $\nu(\lambda, m) = V^{-1}\langle\rho(\lambda)\rangle$. Care must
be taken with the limits; if we reverse the order of the limits so
that we take the chiral limit in a finite box then we will see
symmetry restoration \cite{SHMT}. This is basically due to the fact
that the eigenvalue spectrum has a gap of order $V^{-1}$ when we deal
with a system with a finite number of degrees of freedom (in our case
the instantons).  It is easy to see from equation (\ref{eq:chico})
that the combination of the spectral gap and the absence of any exact
zero modes in full QCD leads to symmetry restoration. This is however
not the case in quenched QCD. In this case we still have the spectral
gap but we also now have order $\sqrt{V}$ exact zero modes, leading to
$\lim_{m \rightarrow 0, V fixed} \ssi \propto 1/m\sqrt{V}$. This
pathological result is well known (see for example \cite{SHMT,ndmt})
and is not the main thrust of the present paper. In this paper we
concentrate on the contribution of the modes which are not
Atiyah-Singer exact zero modes. It is these remaining modes which will
contribute when the limits are taken as in equation
(\ref{eq:chico}). We will therefore ignore the contribution of the
exact zero modes when calculating the chiral condensate. Another
subtlety in equation (\ref{eq:chico}) comes from the explicit quark
mass dependence of the spectral density arising from the presence of
the fermion determinant in the partition function. This quark mass
dependence is absent in the quenched approximation; in this latter
case one may safely evaluate the limits to get

\begin{equation}
\label{eq:B-C}
\ssi = i\pi\nub(0) .
\end{equation}
This is the Banks-Casher relation \cite{bnkcsh}. We immediately see
that the low lying spectral density is crucial for this
phenomenologically important condensate.

Equation~(\ref{eq:B-C}) explains why instantons are a more interesting
starting point for a discussion of chiral symmetry breaking than
perturbation theory. If we neglect interactions then the perturbative
vacuum becomes free. Here the Dirac eigenvalue spectrum grows as
$\lambda^3$. That is to say, the eigenvalues are far from zero. If on
the other hand we neglect interactions amongst (anti)instantons we get
an exact zero-mode for each topological charge. This contributes a
term $\propto \delta(\lambda)$ to $\nub(\lambda)$. Introducing
interactions will shift these distributions, but it is clear that if
what one wants is a non-zero density of modes near $\lambda = 0$ then
the latter approach is a more promising one than the former.

The above discussion suggests the following approach to calculating
the contribution of topological fluctuations to fermionic observables
such as the chiral condensate. First one decomposes the topological
charge density of each gauge field into an ensemble of (overlapping)
instantons and anti-instantons. One then constructs the space spanned
by the would-be zero modes of these instantons.  That is to say, for
each instanton one finds the eigenfunction that would have zero
eigenvalue if that instanton were isolated, and one constructs the
space of linear combinations of these. Within that space one
calculates the required physical observables such as the chiral
condensate. This is most simply done by calculating the eigenvalues
and eigenfunctions of the Dirac operator within this subspace.

This approach already represents an approximation. However to carry it
out requires further approximations. Firstly one needs to identify the
instantons in each gauge field. As remarked earlier we shall use the
instanton ensembles obtained in \cite{DSMT}. To obtain these involved
cooling (smoothing) the lattice gauge fields so as to reveal the
topological charge density and then resolving this into overlapping
(anti)instantons using some `pattern recognition algorithm'. The
topological charge density rapidly changes as we cool a gauge field
but one might reasonably hope that the physically important long
distance fluctuations do not.  For example, the total topological
charge (the $p=0$ projection of the charge density) is very stable
under cooling, if the lattice spacing is small enough. Of course the
total topological charge is a special quantity, and one of our main
aims in this study will be to see how stable is the long distance
fermionic physics under cooling.  The second uncertainty involves the
zero-modes. What we would like to use is a wave-function that
corresponds to the zero mode of an isolated topological charge in the
fully fluctuating vacuum. This is not known and in any case will not
be unique. So we will employ some simple trial wavefunctions which
have the basic property of being localised around the instanton. In
addition we shall neglect the detailed spinorial character of these
wavefunctions except that we keep track of their chirality throughout
the calculations. Clearly one can only believe those results of such a
calculation that are insensitive to the particular form of the
wave-function.  In this paper our calculations will be performed using
a hard-sphere wavefunction. Elsewhere \cite{usprep} we have performed
a detailed comparison with Gaussian wave-functions and with classical
zero modes. The hard-sphere wave-function is particularly simple to
work with in a periodic 4-volume, and produces Dirac spectra similar
to the other trial wave-functions as long as the instanton gas is not
too dilute. As we shall see below, the instanton ensembles that we
take from \cite{DSMT} are typically very dense.  The next
approximation involves the calculation of the eigenvalues of the Dirac
operator. This requires calculating the matrix elements of $\di[A]$
within the space spanned by our zero modes for arbitrary fluctuating
background gauge fields. Since we are not able to do this, we shall
instead take the simplest possible approximation that embodies what we
see as the most essential physics: we replace $\di[A]$ by an operator
which anticommutes with $\gamma_5$ (and thus only has non-zero matrix
elements between states of opposite chirality), which has dimensions
of inverse length as embodied by a factor of $1/l$ where $l$ is a
characteristic length scale of the states appearing in the matrix
element, and which otherwise acts like a unit operator. This series of
approximations retains those qualitative features of the full
calculation that would seem to be the most critical, on the assumption
that the qualitative physics should not depend on all the fine detail
of the fluctuating gauge fields. It may turn out that this assumption
is wrong and that an approach as simple as ours is not possible.

In the next section we introduce in detail our `toy' model for
calculating the contribution of an ensemble of topological charges to
the spectral density. In the following section we apply this model to
the ensembles generated in \cite{DSMT}.

\section{A TOY MODEL FOR EIGENVALUE SPLITTING}

We consider the case of quenched QCD (i.e. the pure gauge theory). We
construct a representation of the Dirac operator \di[A] for a gauge
field A consisting of $n_{I}$ instantons ($I$) and \nni\
anti-instantons ($\overline{I}$). Let each object have gauge field
configuration $A_{i}^{\pm}$. We know that there are $|n_{I} - \nni|$
exact zero modes, whose contribution to the spectral density we will
ignore for the reasons discussed above. There are a further $n_{I} +
\nni - |Q|$ ``would-be zero modes''; these are the modes which would
be exact zero modes if the objects were non-interacting, as in the
dilute gas approximation. In an interacting system these would-be zero
modes become eigenvalues split symmetrically around zero. The
splitting is symmetric as non-zero eigenvalues come in pairs
$\pm\lambda$ due to the $\gamma^{5}$ symmetry $\{\di,\gamma^{5}\} =
0$. It is these ``would-be zero modes'' which we expect to form that
part of the low lying spectrum of the Dirac operator that is due to
instantons, and so are of greatest importance in calculations of the
chiral condensate.

A matrix representation is defined by $(\di[A])|j\rangle =
D_{kj}|k\rangle$ for some basis $\{|j\rangle\}$. It follows trivially
that $D_{kj} = \langle k|\di|j\rangle$ iff the basis is
orthonormal. We choose our basis to be constructed out of the would-be
zero modes coming from the individual objects
$\{|\psi_{1}^{+}\rangle,\ldots,|\psi_{\nni}^{+}\rangle,|\psi_{1}^{-}\rangle,\ldots,|\psi_{n_{I}}^{-}\rangle\}$
such that

\begin{equation}
\label{eq:mot}
\di[A_{i}^{\pm}]|\psi_{i}^{\pm}\rangle = 0 .
\end{equation}
We are in effect writing the low lying eigenmodes of a given instanton
configuration as a linear combination of the zero modes from the
individual objects. If we were interested in the entire spectrum then
we would have to include contributions from all modes but this is not
necessary for our purpose. It is easy to see that

\begin{eqnarray}
\langle\psi_{i}^{+}|\di[A]|\psi_{j}^{+}\rangle & = & 0\nonumber\\
\langle\psi_{i}^{-}|\di[A]|\psi_{j}^{-}\rangle & = & 0\nonumber\\
\langle\psi_{i}^{+}|\di[A]|\psi_{j}^{-}\rangle & \doteq &
V_{ij} .
\end{eqnarray}
The first two identities follow from the $\gamma^{5}$ symmetry. Let us
consider the third equation for a configuration consisting of a single
$I\!-\!\overline{I}$ pair. The function $V$ has, in principle, a
dependence on the position of the centres of the objects
($x_{k}^{\pm}$), the sizes of the objects ($\rho_{k}^{\pm})$ and the
relative colour orientation of the two objects. We choose to ignore
the colour orientation for reasons of simplicity. We can also replace
the covariant derivative $\di[A]$ in the matrix element with
$-i\!\!\not\!\partial$ using the equations of motion
(\ref{eq:mot}). So given a trial would-be zero mode wavefunction, we
can actually calculate this matrix element as an overlap integral over
some manifold (in all cases in this paper the manifold is simply the
periodic box ${\mathbb T}^{4}$).

\begin{eqnarray}
\label{eq:whatisV}
V_{ij} & = &
\langle\psi_{i}^{+}|\di[A_{i}^{+},A_{j}^{-}]|\psi_{j}^{-}\rangle\nonumber\\
& = & \langle\psi_{i}^{+}|-i\!\!\not\!\partial|\psi_{j}^{-}\rangle\nonumber\\
& = & V(x_{i}^{+},\rho_{i}^{+},x_{j}^{-},\rho_{j}^{-}) .
\end{eqnarray}
Here we have assumed that the $I\!-\!\overline{I}$ gauge potential can
be approximated as a sum of the individual $I,\overline{I}$ gauge
potentials (in some appropriate singular gauge). We note that
(\ref{eq:whatisV}) will be valid for an arbitrary number of
(anti)instantons if their mutual separations are large. If however we
are dealing with an arbitrary configuration consisting of more than a
single $I\!-\!\overline{I}$ pair, then the function V is in general
unknown. It will in fact have a dependence on all the objects within
the configuration through the gauge field $[A]$. We also cannot
replace the covariant derivative as before. We however choose to make
this replacement anyway so that (\ref{eq:whatisV}) holds for general
configurations. Again, the main justification for doing so is
simplicity.

In terms of this basis (and noting the above reservations), we can
construct the $(n_{I} + \nni)\times(n_{I} + \nni)$ matrix $D \equiv
\langle\psi|\di|\psi\rangle$ which has block zeroes on the diagonal
and $V, V^{\dagger}$ off block diagonal. There is a however a
fundamental objection to $D$ being thought of as a matrix
representation of the Dirac operator $\di[A]$. This is due to the fact
that the basis we have chosen is not orthonormal;

\begin{eqnarray}
\langle\psi_{i}^{+}|\psi_{j}^{+}\rangle & = & U(x_{i}^{+},\rho_{i}^{+},x_{j}^{+},\rho_{j}^{+})\nonumber\\
\langle\psi_{i}^{-}|\psi_{j}^{-}\rangle & = & U(x_{i}^{-},\rho_{i}^{-},x_{j}^{-},\rho_{j}^{-})\nonumber\\
\langle\psi_{i}^{\pm}|\psi_{j}^{\mp}\rangle & = & 0 ,
\end{eqnarray}
hence the matrix $D$ is simply not a representation. (We have again
ignored the possible dependence of the matrix element on the relative
colour orientation of the objects; this is again for simplicity.) We
can however construct an orthonormal basis
$\{|\widetilde{\psi}^{+}\rangle,|\widetilde{\psi}^{-}\rangle\}$ using
the standard Gram-Schmidt procedure,

\begin{eqnarray}
|\psi_{j}^{+}\rangle & = & R_{ij}|\widetilde{\psi}_{i}^{+}\rangle\quad 1 \leq
i \leq j \leq \nni\nonumber\\
|\psi_{j}^{-}\rangle & = & S_{ij}|\widetilde{\psi}_{i}^{-}\rangle\quad 1 \leq
i \leq j \leq n_{I} .
\end{eqnarray}
It is easy to see that in terms of this new orthonormal basis, a
matrix representation of the Dirac operator is given by:

\begin{equation}
\begin{array}{ccc}
\hspace{1.8cm}\overbrace{\hspace{2.7cm}}^{n_{\overline{I}}} & \hspace{0.0cm}\overbrace{\hspace{2.7cm}}^{n_{I}} & \\
\di \doteq \widetilde{D} = 
\left(
\begin{array}{l}
\langle\widetilde{\psi}_{i}^{+}|\di|\widetilde{\psi}_{j}^{+}\rangle = 0 \\
\langle\widetilde{\psi}_{i}^{-}|\di|\widetilde{\psi}_{j}^{+}\rangle = \widetilde{V}_{ij}^{\dagger}
\end{array}
\right. & 
\left.
\begin{array}{l}
\langle\widetilde{\psi}_{i}^{+}|\di|\widetilde{\psi}_{j}^{-}\rangle =
 \widetilde{V}_{ij}\\
\langle\widetilde{\psi}_{i}^{-}|\di|\widetilde{\psi}_{j}^{-}\rangle = 0
\end{array}
\right) &
\begin{array}{l}
\Big\}\ \nni\\ \Big\}\ n_{I}
\end{array}
\end{array}
\label{eq:dirm}
\end{equation}
where
\begin{eqnarray}
\widetilde{V} & = & (R^{-1})^{\dagger}VS^{-1}
\end{eqnarray}
We see that this orthonormalization procedure preserves the chiral
properties of the original wavefunctions;
$\gamma^{5}|\widetilde{\psi}^{\pm}\rangle =
\pm|\widetilde{\psi}^{\pm}\rangle$. This matrix representation
fulfills many of the requirements for the Dirac operator. It is easy
to see that the elements of the matrices $R, S$ and $D$ are not
random; they obey various triangle inequalities as they are given by
overlap integrals between wavefunctions. Gram-Schmidt
orthonormalization would be impossible for a general matrix without
this property. It is also easy to see that $\widetilde{D}$ satisfies
the $\gamma^{5}$ symmetry; all non-zero eigenvalues come in pairs
$\pm\lambda$. Furthermore we see that the Atiyah-Singer theorem is
obeyed; any configuration has at least $|n_{I} - \nni|$ exact zero
eigenvalues. If we consider the case of a single $I\!-\!\overline{I}$
pair (so orthonormalization becomes trivial), then it is easy to see
that by choosing the function $V$ appropriately, we can recover the
correct eigenvalue splitting. (This is given by the overlap matrix
element between would-be zero mode wavefunctions.) All these features
give us reason to believe that our toy model should capture the
essential features of the mechanism whereby eigenvalues are split from
zero.

All we have to decide is what form of trial would-be zero mode
wavefunction to use for our objects. Simple examples include the hard
sphere, Gaussian and classical zero mode wavefunctions. It is of
course not possible to say which wavefunction will dominate the
quantum vacuum. In this paper we concentrate on the simplest possible
case, that of hard sphere wavefunctions;

\begin{eqnarray}
\langle x|\psi_{j}^{\pm}\rangle & = & 1\qquad |x - x_{j}^{\pm}| \leq
\rho_{j}^{\pm}\nonumber\\
& = & 0\qquad \rm{otherwise} .
\end{eqnarray}
This choice would of course lead to an artificially singular form for
$V_{ij}$ and so (again in the interests of simplicity) we replace the
derivative with $1/\sqrt{\rho_{1}\rho_{2}}$ so that both U and V are
modeled by hard sphere wavefunctions but with the correct dimensions.
We leave it to another publication to investigate more closely the
universality of our results \cite{usprep} for different wavefunction
choices. In that study we perform extensive comparisons between
hard-sphere, Gaussian and classical zero-mode wave functions, both on
${\mathbb R}^{4}$ and on ${\mathbb T}^{4}$ for a range of packing
fractions and volumes. Although incomplete, this study shows that the
qualitative features emphasised in the present paper are indeed
universal. It is only these features, which are largely independent of
the trial wavefunction used, that can be considered to be of
significance and we can hope will survive in the full quenched QCD
vacuum.

Each configuration in the ensemble consists of a set of positions,
sizes and winding numbers ($\pm 1$) which label the objects in the
vacuum. This set is either extracted from gauge field configurations
generated by UKQCD \cite{DSMT} or alternatively, produced by a model
that simply generates them at random. The advantage of the former
method is that it may give a more accurate description of the
topological content of the quenched QCD vacuum; the latter method has
the advantage that we can obtain far higher statistics. And in some
sense it isolates more cleanly the physics due to instantons, since
the instantons in the lattice fields certainly contain correlations
which are due to other dynamics.  We use the method outlined above to
find the low lying eigenvalues for each given configuration. These are
then used to generate the spectral density.

\section{RESULTS}

In \cite{DSMT} SU(3) lattice gauge field configurations of sizes $16^3
48$ at $\beta=6.0$, $24^3 48$ at $\beta=6.2$ and $32^3 64$ at
$\beta=6.4$ were cooled and the corresponding instanton ensembles
extracted for various numbers of cooling sweeps. (We shall in the
following frequently use ``instantons'' as a short-hand for
``instantons and anti-instantons''.) Over this range of $\beta=6/g^2$
the lattice spacing varies by a little less than a factor of 2 and
these three volumes are approximately the same in physical
units. Comparing the results at the three values of $\beta$ enables
the approach to the continuum limit to be studied.  Of course,
instantons can be large and it is important to control finite volume
effects as well. For this purpose calculations were also performed at
$\beta=6.0$ on a much larger $32^3 64$ lattice. The conclusion was
that finite volume corrections were negligible and that there was good
scaling of, for example, the instanton size distribution, if one
varied the number of cooling sweeps with $\beta$ so as to keep the
average number of instantons constant. (For an interesting recent
analysis of the scaling properties, see \cite{arfs}.)  Some properties
of these lattice configurations are listed in Table~\ref{tab:cdata}.

In this section we shall take these configurations, ranging in number
from 20 to 100 depending on the lattice size and the value of $\beta$,
and we calculate the spectral density of the would-be zero modes as
described in the previous section. We shall, for simplicity, not
employ some of the rather complicated procedures used in \cite{DSMT}
for filtering out possible false instanton assignments. Rather we
shall take the raw instanton ensembles from \cite{DSMT}, corrected for
the influence of the instantons upon each other but without applying
any further filters. (Except that we throw away any charges that are
larger than the volume available.  This usually involves rejecting
(much) less than $1\%$ of the total number.) In addition, we calculate
the size from the (corrected) peak height. We are confident that the
results we obtain from these ensembles differ very little from the
results we would have obtained using the slightly different ensembles
obtained by applying the more complex procedures of \cite{DSMT}.

There are several questions we wish to address. These include:

{\noindent}$\bullet$ Do fermionic physical observables, such as the
spectral density and the chiral condensate, exhibit a weak variation
with cooling, implying that the rapid variation of the instanton
ensemble that one observes is more apparent than real, or do they
exhibit a strong variation?

{\noindent}$\bullet$ Do these fermionic physical observables also
exhibit scaling and small finite volume corrections?

{\noindent}And more generally:

{\noindent}$\bullet$ Does a realistic ensemble of instantons break
chiral symmetry spontaneously? Lattice calculations find that it does;
but the presence of important lattice artefacts renders the conclusion
suspect. Continuum calculations using model ensembles of instantons
also find that they break chiral symmetry; but it is not clear that
the real world is like the model.

{\noindent}$\bullet$ Is the spectral density of quenched QCD
pathological? Some model calculations have found that the spectrum
appears to diverges at $\lambda=0$ \cite{ndmt,usmt,dvgce-other}.
Whether this is related to the appearance, within quenched chiral
perturbation theory, of a logarithmic \cite{dvgce-other,cpt-log}, or
possibly power divergence \cite{cpt-power}, is also of interest.

\begin{center}
\begin{figure}[p]
\leavevmode
\epsfxsize=100mm
\epsfbox{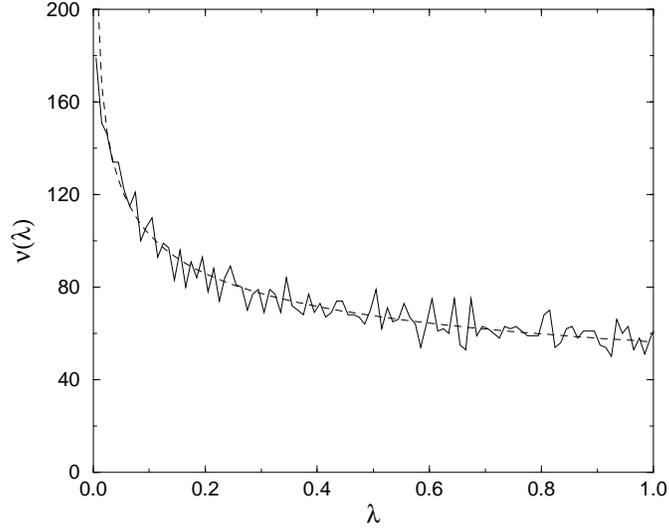}
\caption{The spectral density: $\beta = 6.0,\ 32^{3}64,\ 50$
configurations. Dashed curve is best power law fit $\nu(\lambda) = a
+ b/\lambda^{d}$.}
\label{fig:0.46L}
\end{figure}
\end{center}

\begin{center}
\begin{figure}[p]
\leavevmode
\epsfxsize=100mm
\epsfbox{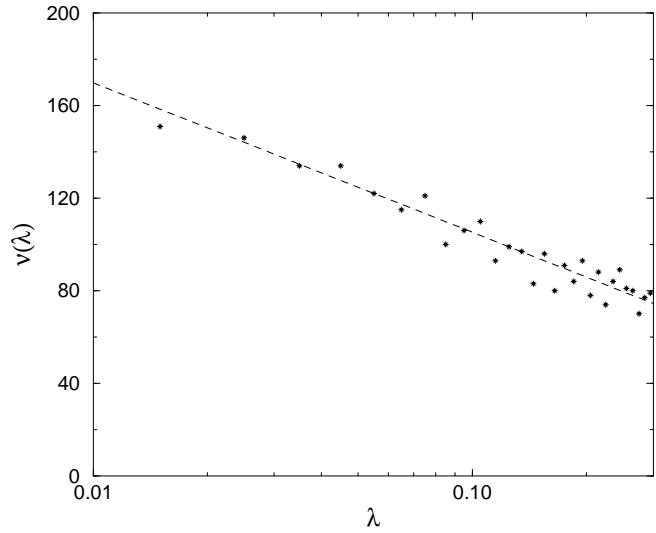}
\caption{The spectral density in figure~\ref{fig:0.46L}, 
plotted on log-linear axes. Dashed curve is best log law fit
$\nu(\lambda) = a + b\ln(\lambda)$.}
\label{fig:0.46L.lnlin}
\end{figure}
\end{center}

\begin{center}
\begin{figure}[p]
\leavevmode
\epsfxsize=100mm
\epsfbox{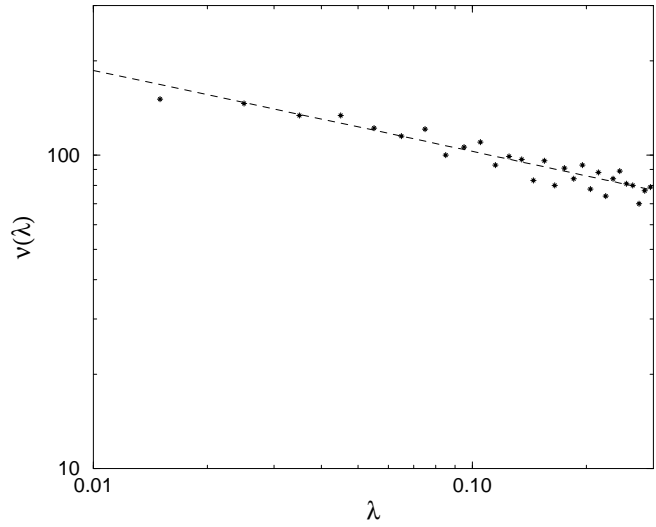}
\caption{The spectral density in figure~\ref{fig:0.46L}, 
plotted on log-log axes. Dashed curve is best power law fit
$\nu(\lambda) = a + b/\lambda^{d}$.}
\label{fig:0.46L.lnln}
\end{figure}
\end{center}

\begin{center}
\begin{figure}[p]
\leavevmode
\epsfxsize=100mm
\epsfbox{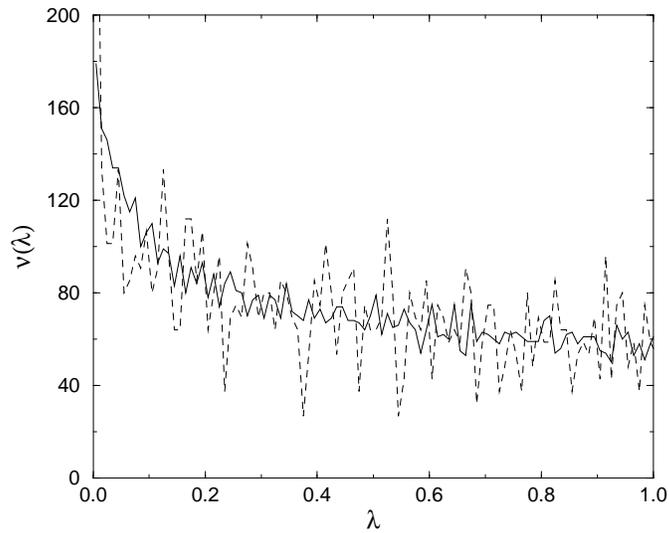}
\caption{Spectral densities from two different volumes
at $\beta=6.0$, after 46 cooling sweeps:
$32^3 64$(solid) and $16^3 48$(dashed) lattices.}
\label{fig:0.46L.vol}
\end{figure}
\end{center}

\begin{center}
\begin{figure}[p]
\leavevmode
\epsfxsize=100mm
\epsfbox{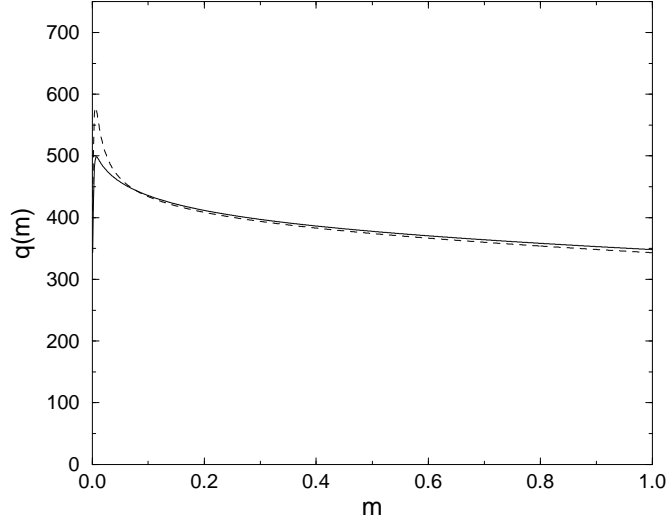}
\caption{\ssi = $(q(m) \rm{MeV})^{3}$ as obtained from the $\beta =
6.0\ 32^{3}64$ lattice data (solid), and from the $\beta = 6.0\
16^{3}48$ lattice data (dashed).}
\label{fig:0.46x.qc}
\end{figure}
\end{center}

Figure~\ref{fig:0.46L} shows the spectral density that results from
the 50 configurations generated after 46 cooling sweeps on the
$32^{3}64$ lattice at $\beta = 6.0$. We see that the spectral density
does not smoothly decrease to zero as $\lambda \rightarrow 0$, so the
chiral symmetry will be spontaneously broken.  However we also see a
pronounced peak as $\lambda \rightarrow 0$, just like the divergence
that characterises model instanton ensembles \cite{ndmt,usmt}.  (Note
that the exact zero modes contribute an invisible
$\delta$-function. Because this should be regarded as a finite volume
effect, we shall not include it in the calculations of this section.)
What is the functional form of this peak ? We attempt to model the
peak as a log divergence $\nu(\lambda) = a + b\ln(\lambda)$ and as a
power divergence $\nu(\lambda) = a + b/\lambda^{d}$.
Figures~\ref{fig:0.46L.lnlin} and \ref{fig:0.46L.lnln} show that both
models fit the data well, an observation which is confirmed by the low
chi-squared for the fit (see Table~\ref{tab:cdata}).  (The error
estimates use the `jack-knife' method and correlations between the
errors in different eigenvalue bins are neglected.  The observed
scatter of values suggests that the latter should not be too bad an
approximation.) We should check that this result is not subject to
large finite volume effects and to this end we compare the spectral
density to that obtained from the $\beta = 6.0, 16^{3}48$ lattice (a
volume approximately ten times smaller). As shown in
figure~\ref{fig:0.46L.vol}, whilst the result on the smaller volume is
noisier, we find the densities are entirely similar, even down to the
details of the forward peak. This shows that at least for these
parameters any finite volume corrections are small. The chiral
condensate as a function of quark mass is given in
figure~\ref{fig:0.46x.qc}. We know that this order parameter must
vanish for very small quark masses because of the gap in the
eigenvalue spectrum (we cannot take the quark mass to zero in a finite
box) and indeed it does.  If we extrapolate to zero quark mass, whilst
ignoring the finite volume dip at very small quark masses and the peak
at small masses, we find that chiral symmetry is broken with an order
parameter $\ssi^\frac{1}{3} \approx 400 {\rm MeV}$. Whilst this is
larger than the phenomenological figure, $\ssi^\frac{1}{3} \sim 200
{\rm MeV}$, it is close considering the qualitative nature of our
calculations. The fact it is larger is presumably a reflection of the
high density of this gas of instantons.

Although the qualitative features of our spectrum do not require a
specification of units, the comparison between different instanton
ensembles does. The units we have chosen are as follows. Our length
unit is chosen to be $32a$ at $\beta=6.0$; so that the $32^3 64$ and
$16^3 48$ lattices discussed in the previous paragraph have volumes 2
and 0.1875 respectively.  We see from Table~\ref{tab:cdata} that this
corresponds to taking our length unit as $32a(\beta=6.0) \simeq 32
\times 0.098fm = 3.136 fm$.  Thus our mass unit is the inverse of
this, $\simeq 64 MeV$.  Since $\lambda$ has dimensions $[m]^1$, this
means that the eigenvalues shown in Figure~\ref{fig:0.46L} range from
0 to $\simeq 64 MeV$: a reasonable range if what we are interested in
is the spectrum $\lambda < \Lambda_{QCD}$. We maintain this unit
throughout the calculations of this paper and we use the values of
$a(\beta)$ listed in Table~\ref{tab:cdata} to translate this unit to
other values of $\beta$. Thus if we want to test for scaling all we
need to do is to directly superpose the spectra as shown in our
figures. (We shall do this later on in this section.)

\begin{center}
\begin{figure}[p]
\leavevmode
\epsfxsize=100mm
\epsfbox{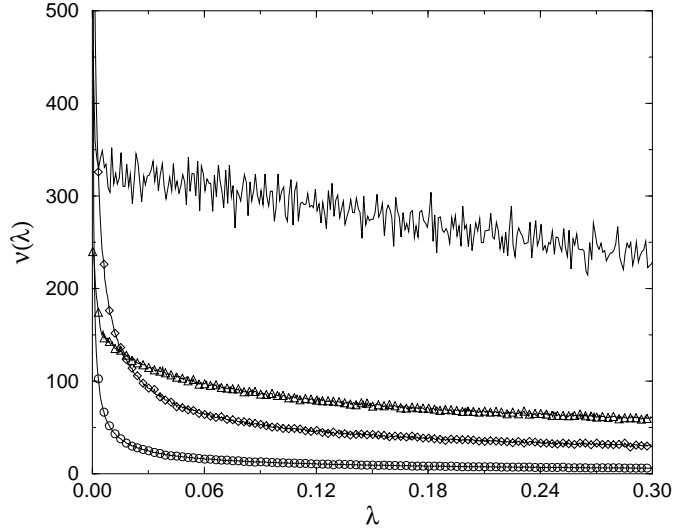}
\caption{Spectral densities from configurations generated by the
random-position model, for various packing fractions, keeping
the volume fixed and varying the number of charges as shown:
$N\overline{V}_{I}/L^{4}$ =
0.2 ($\circ\ N = 13 + \overline{13}$), 1.0 ($\diamond\ N = 63 +
\overline{63}$), 2.5 ($\triangle\ N = 153 + \overline{153}$) \& 10.0
 (solid $N = 633 + 633$).}
\label{fig:h_xxx_020}
\end{figure}
\end{center}

\begin{center}
\begin{figure}[p]
\leavevmode
\epsfxsize=100mm
\epsfbox{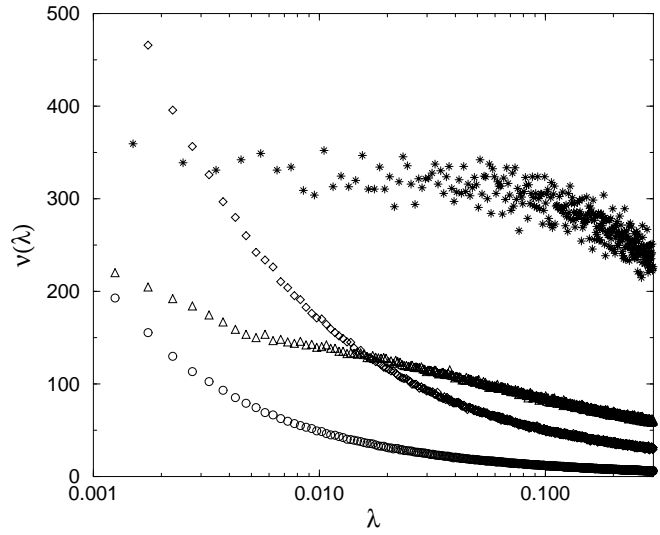}
\caption{Spectral densities in figure (\ref{fig:h_xxx_020}), 
plotted on log-linear axes.}
\label{fig:h_xxx_020_lnlin}
\end{figure}
\end{center}

\begin{center}
\begin{figure}[p]
\leavevmode
\epsfxsize=100mm
\epsfbox{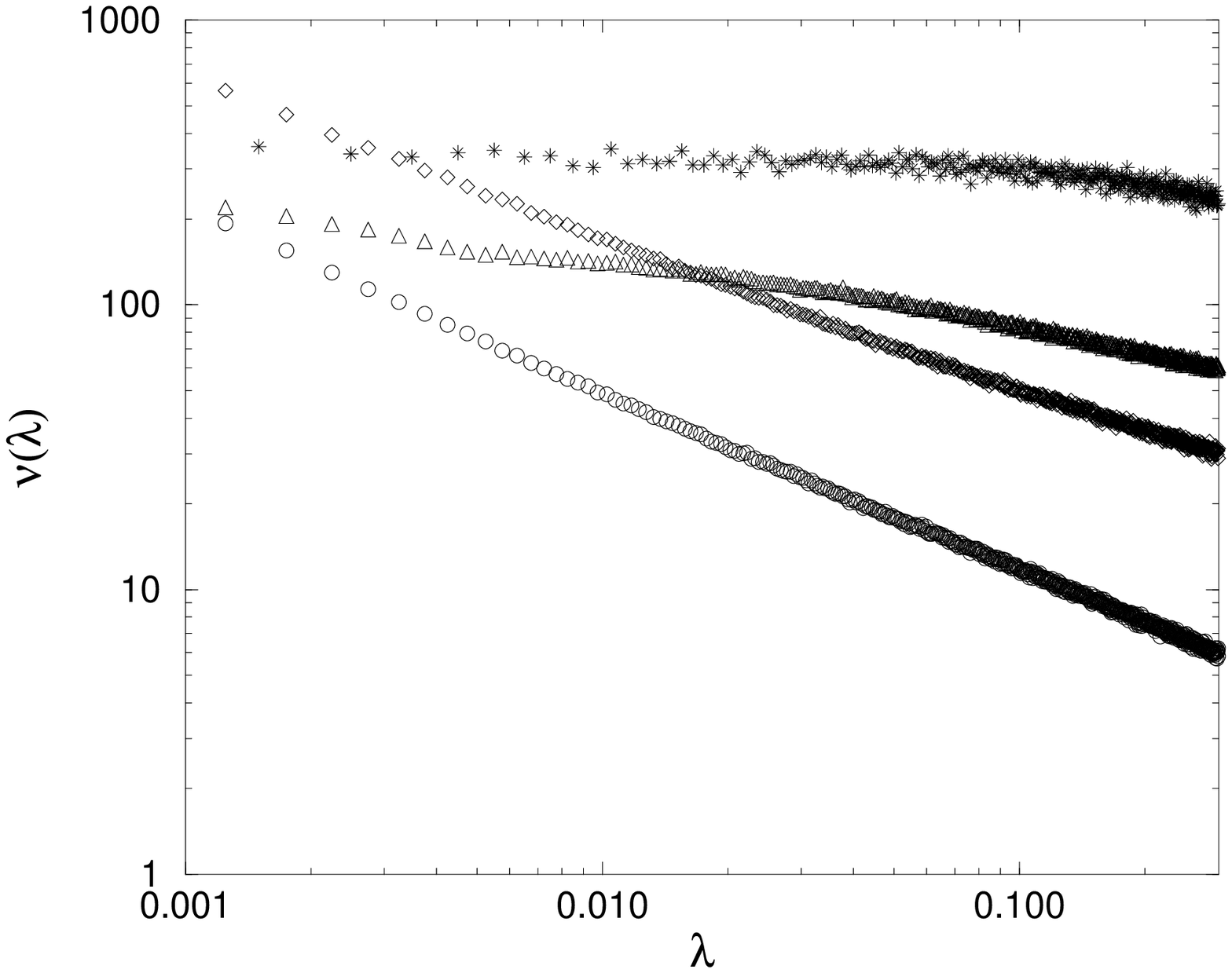}
\caption{Spectral densities in figure (\ref{fig:h_xxx_020}), 
plotted on log-log axes.}
\label{fig:h_xxx_020_lnln}
\end{figure}
\end{center}

\begin{center}
\begin{figure}[p]
\leavevmode
\epsfxsize=100mm
\epsfbox{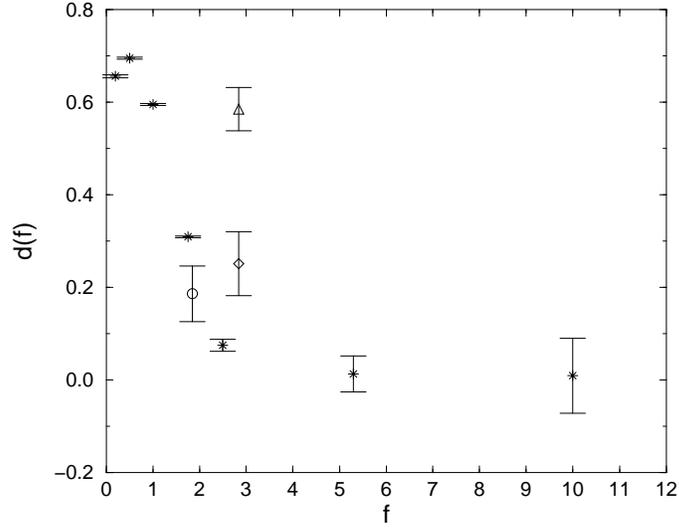}
\caption{Plot of power of divergence $d$ as a function of packing
fraction $f$. $\star$ from random position model. $\diamond$ from
lattice data as in figure (\ref{fig:0.46L}). $\circ$ same, except all
instantons of same size $\overline{\rho}$. $\triangle$ as figure
(\ref{fig:0.46L}) except instantons positioned at random.}
\label{fig:DvsF}
\end{figure}
\end{center}

We have focussed on the lattice calculations at $\beta=6.0$ and after
46 cooling sweeps because it is only here that we can perform an
explicit finite volume study. To go beyond this constraint, and that
imposed by the modest statistical accuracy of the lattice data, it is
useful to analyse instanton configurations that have been generated in
a simple toy model where the topological charges are placed at random
positions in spacetime \cite{usprep,ndmt,usmt}.  All these
configurations are made to have zero topological charge since we know
that $Q/V \to 0$ as $V\to\infty$, so we expect that taking only $Q=0$
minimises finite volume corrections. A further simplification is that
all charges in the random position model are the same size. By
comparing this with the lattice data, where the instantons have
various sizes, we will be able to investigate the effects of allowing
the instanton sizes to vary. In figure~\ref{fig:h_xxx_020} we plot a
few of the spectral densities obtained from these ensembles of
synthetic configurations. The different spectral densities are
obtained by varying the packing fraction of the configurations
(holding other parameters constant). The packing fraction is defined
as the average number of topological charges multiplied by their
average volume and divided by the total volume of spacetime, i.e. the
quantity $\overline{N}_{I}\overline{V}_{I}/V$ in
Table~\ref{tab:cdata}.  It is a measure of the fraction of the
spacetime that would be occupied if the objects were non-overlapping:
increasing the packing fraction whilst holding other variables
constant increases the density of objects. We vary the packing
fraction from 0.2 to 10 for high statistics ensembles (a minimum of
1266 configurations for the highest density to over 650000
configurations for the lowest). We observe a strong $\lambda\to 0$
divergence at low packing fractions which weakens as the density
increases. As far as the functional form of this peak is concerned,
Figure~\ref{fig:h_xxx_020_lnlin} and table~\ref{tab:sdata} show that
we can rule out the log divergence as a model for the peak for low
density ensembles (packing fractions less than approximately 2). The
power law divergence however gives a good fit for all densities (see
figure~\ref{fig:h_xxx_020_lnln} and table~\ref{tab:sdata}). We note
that it is only when we approach dense gases that the logarithmic fit
to the divergence works equally well. Presumably this is simply a
trivial consequence of the weakness of the power divergence, since
\begin{equation}
\label{eq:log-power}
b\lambda^{-d} \simeq b - bd\ln(\lambda)
\end{equation}
for small values of the exponent $d$.  We plot a graph of the degree
of divergence (the power $d$ associated with the $\lambda^{-d}$
divergence) versus the packing fraction in figure~\ref{fig:DvsF}. This
shows that the divergence weakens as the packing fraction
increases. It is not clear within our accuracy, whether there is a
non-zero albeit very small divergence at very high densities, or
whether the degree of divergence vanishes at some finite density. For
practical purposes this question is not important; once the divergence
is weak enough it will in any case have a negligible effect on the
chiral physics in the vicinity of the physical quark masses.  All this
fits well with the naive picture whereby a higher packing fraction
leads to greater overlaps (on average), leading to greater splitting
of eigenvalues from zero, hence a weaker divergence (if any).

\begin{center}
\begin{figure}[p]
\leavevmode
\epsfxsize=100mm
\epsfbox{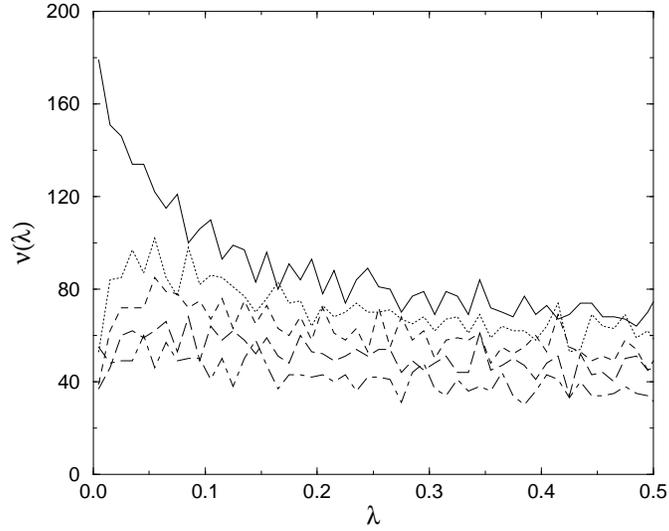}
\caption{Lattice Data: Solid line as figure (\ref{fig:0.46L}). Dotted
line only contains objects with $\rho > 0.12$, dashed line $\rho >
0.14$, long dashed $\rho > 0.16$, dot-dash $\rho > \overline{\rho} \approx
0.182$.}
\label{fig:gt0xx}
\end{figure}
\end{center}

In figure~\ref{fig:DvsF} we also plot the degree of divergence
corresponding to the $\beta = 6.0, 32^{3}64$ lattice data. It would
appear that the divergence is too large for the packing fraction; it
lies approximately two sigma above the synthetic ensemble curve. It is
therefore interesting to consider which particular aspect of the
lattice data contributes to the divergence. The first possible factor
is the non-trivial size distribution of the objects associated with
the lattice data. We therefore set all objects in the lattice data to
the same size (the mean $\overline{\rho}$ of the lattice
ensemble). This reduces the packing fraction of the ensemble, since
$\overline{\rho}^4 < \overline{\rho^4}$, and, as shown in
figure~\ref{fig:DvsF}, it also results in the degree of divergence
fitting with that of the synthetic ensemble.  The fact that a
non-trivial instanton size distribution has a marked impact on the
spectrum of small modes leads us to ask whether it is the small or the
large instantons that drive this effect. To answer this question we
systematically cull instantons of ever increasing size from the
lattice instanton ensembles and see how this affects the spectral
density.  The results of this calculation are shown in
figure~\ref{fig:gt0xx}.  In this figure we show the densities obtained
by only including objects with radii above a certain cut-off. We see
that the peak is already significantly reduced if we exclude the $\sim
10\%$ of instantons with radii $\rho < 0.12$; and it is eliminated
entirely if we exclude all instantons with $\rho < \bar{\rho} \simeq
0.18$.  (If on the other hand we exclude the largest instantons, then
we find that we strengthen the peaking at $\lambda=0$.)  This shows
that the extra peaking we have observed with the lattice instanton
ensembles is due to the smaller instantons.  More generally, this
demonstrates that it is possible to have small instantons driving a
`divergent' spectral density even in a high density gas. The reason
for this unexpected phenomenon is actually quite simple. The large
packing fraction of such a gas is driven by the larger instantons
(since the volume $\propto \rho^4$).  The smaller instantons are
rather dilute and are not likely to overlap significantly with each
other. Instead they typically overlap completely with some of the much
larger instantons.  However this overlap is small: if we have a small
instanton of radius $\rho_{s}$ sitting on a large one of size
$\rho_{l}$ (of the opposite charge) then this will contribute $\propto
\rho_{s}^{4}/\rho_{s}^{2}\rho_{l}^{2} = (\rho_{s}/\rho_{l})^{2}$ to
the overlap matrix.  The larger instantons, on the other hand, will
have large overlaps with other large instantons (of the opposite
charge) in addition to their small overlaps with small instantons.  So
they are less likely candidates for producing small eigenvalues. Thus
the smaller instantons in an apparently dense gas can behave as a
dilute gas with a corresponding peak at small eigenvalues. This would
seem to be a phenomenon that might well be peculiar to quenched rather
than full QCD.

\begin{center}
\begin{figure}[p]
\leavevmode
\epsfxsize=100mm
\epsfbox{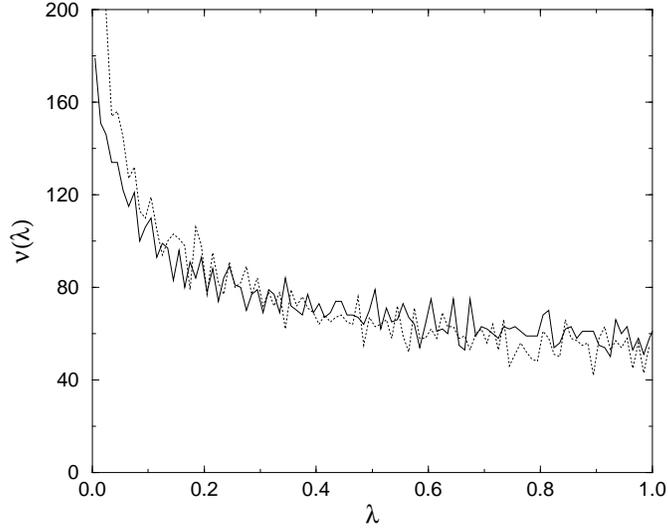}
\caption{The spectral density of figure~\ref{fig:0.46L} (solid), 
compared to the density obtained by positioning the same
charges at random (dotted).}
\label{fig:0.46L.rpos}
\end{figure}
\end{center}

It is also interesting to ask whether the non-random positioning of
the instantons in the lattice ensembles makes a difference to the
small-$\lambda$ peak in the spectral density. We see from
figure~\ref{fig:DvsF} and figure~\ref{fig:0.46L.rpos} that it does;
positioning the objects at random (but incorporating other information
such as the size distibution) increases the degree of
divergence. Whilst there are systematic uncertainties (due to deciding
which region of the data to fit the model to), this result is seen in
all the lattice data that we have analysed. The simplest explanation
for this is that we are seeing an effect of the topological charge
screening that was observed in \cite{DSMT}.  This tendency for
opposite charges to `pair up', will lead to an increased eigenvalue
splitting and a weaker divergence. When we position the objects at
random, this screening is lost and the degree of divergence is
increased.

\begin{center}
\begin{figure}[p]
\leavevmode
\epsfxsize=100mm
\epsfbox{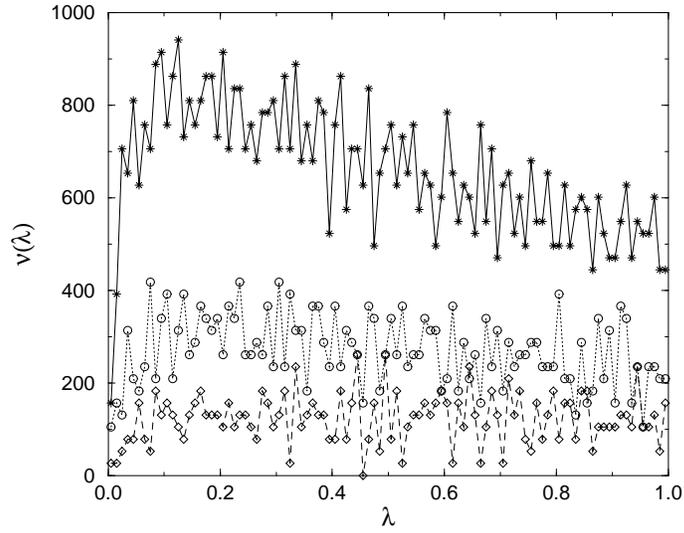}
\caption{The spectral densities obtained from the 
$\beta = 6.4,\ 32^{3}64$ configurations for various numbers of 
cooling sweeps: ($\star$) 30 cools, ($\circ$) 50 cools, ($\diamond$)
80 cools.}
\label{fig:4.xx.sp}
\end{figure}
\end{center}

\begin{center}
\begin{figure}[p]
\leavevmode
\epsfxsize=100mm
\epsfbox{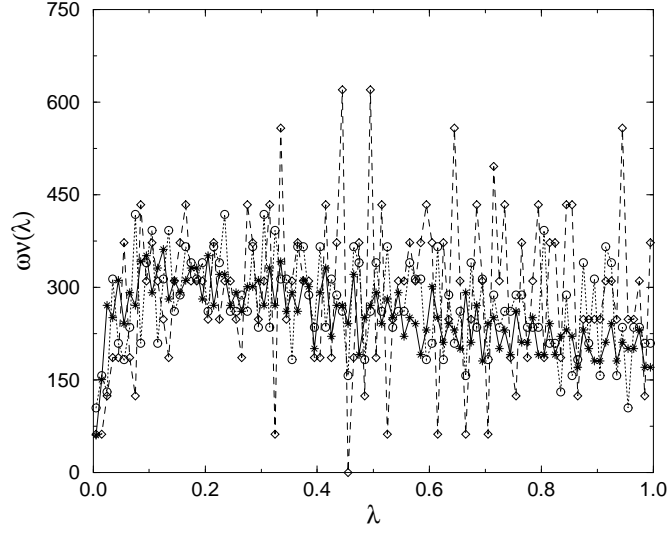}
\caption{The same densities as in figure~\ref{fig:4.xx.sp} but 
rescaled by $\omega = n_{50}/n_{i}$, where $n_{i}$ is number of 
topological charges after $i$ cooling sweeps.}
\label{fig:4.xx.sp.res}
\end{figure}
\end{center}

\begin{center}
\begin{figure}[p]
\leavevmode
\epsfxsize=100mm
\epsfbox{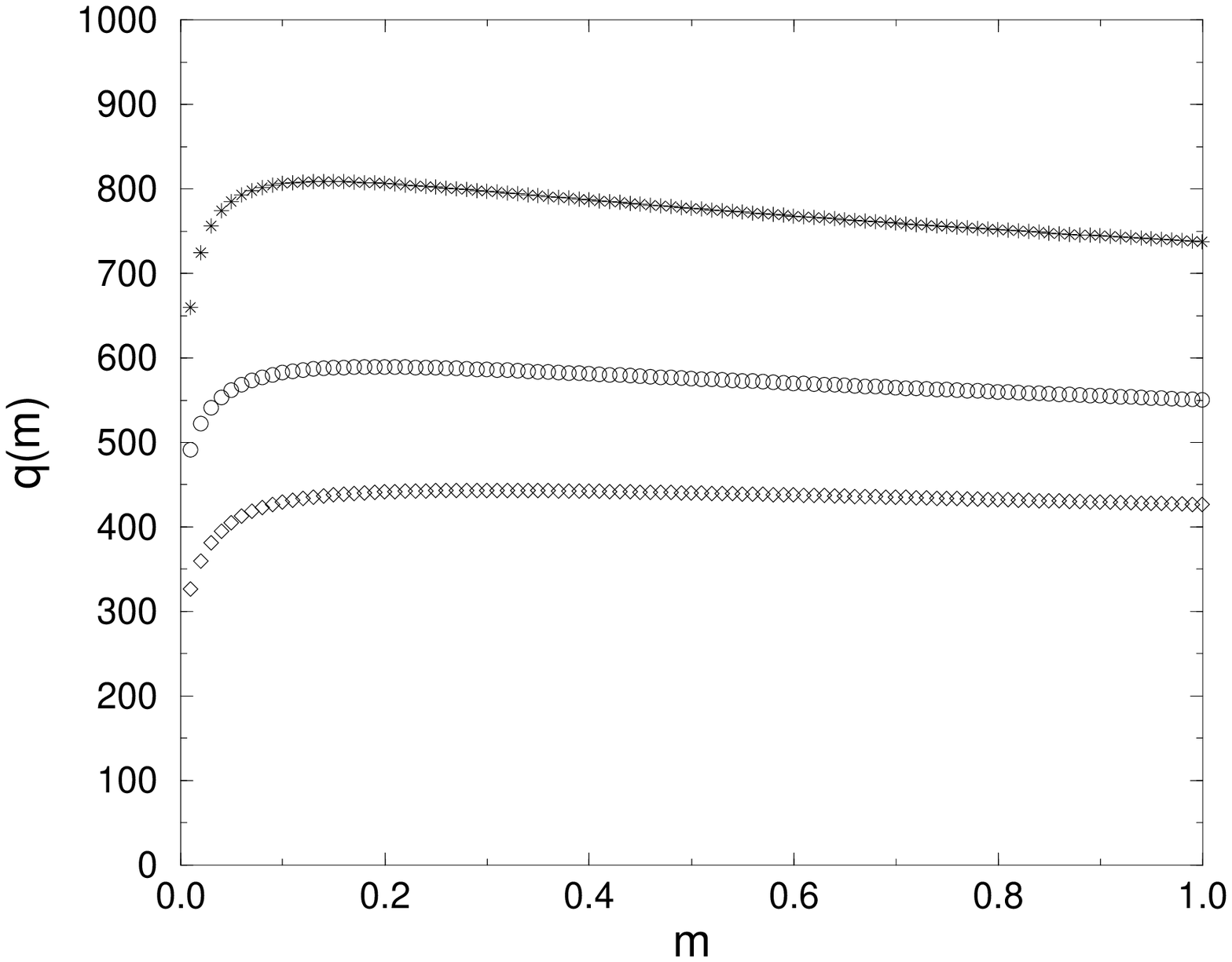}
\caption{\ssi = $(q(m) \rm{MeV})^{3}$
 obtained from the spectral densities in
figure~\ref{fig:4.xx.sp}.} 
\label{fig:4.xx.qc}
\end{figure}
\end{center}

\begin{center}
\begin{figure}[p]
\leavevmode
\epsfxsize=100mm
\epsfbox{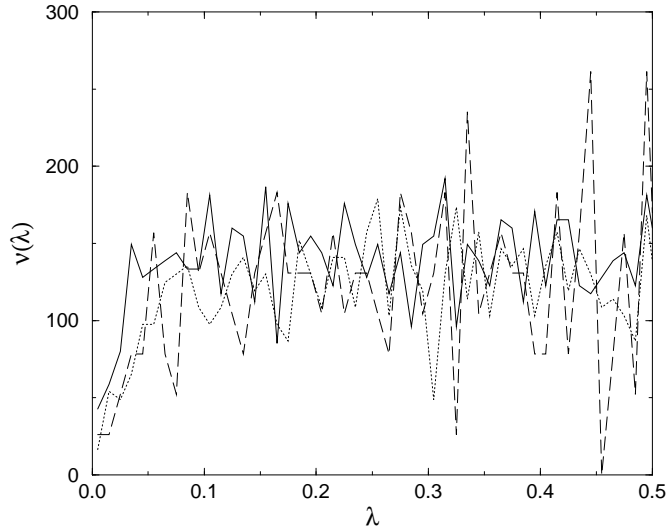}
\caption{The spectral densities obtained at $\beta=6.0, \ 6.2, \ 6.4$
after 23 (solid), 46 (dotted) and 80 (dashed) cooling sweeps respectively.}
\label{fig:x.xx.scl}
\end{figure}
\end{center}

We return now to analyzing the lattice data and specifically to
testing the effects of cooling. As a lattice field configuration is
cooled, one finds \cite{DSMT} that the average size of the instantons
increases and opposite charges annihilate. The former leads to fewer
smaller objects whilst the latter leads simply to fewer objects in
total per unit volume. Figure~\ref{fig:4.xx.sp} shows the spectral
density for the 20 configurations generated at $\beta = 6.4$
(corresponding to the smallest lattice spacing) on a $32^{3}64$
lattice for 30, 50 and 80 cooling sweeps respectively. The
configurations after 80 sweeps are thought to correspond to
configurations after 23 sweeps at $\beta = 6.0$ (see \cite{DSMT} and
figure~\ref{fig:x.xx.scl}). Hence all these configurations are denser
than those analysed previously. We might therefore expect the peaking
at $\lambda=0$ to be weaker, or even non-existent. This is indeed what
we see in figure~\ref{fig:4.xx.sp}. We also see something rather
striking; as we cool more, and as we find fewer objects in the same
volume, the entire spectral density shifts downwards in a way that is
roughly proportional to the change in instanton number (see
figure~\ref{fig:4.xx.sp.res}).  This is in contradiction with the
optimistic expectation that cooling, being a local smoothing, should
have less effect at small eigenvalues (`infrared physics') and more
effect at large eigenvalues (`ultraviolet physics') -- as would occur
if the main reason for the decrease of the number of charges with
cooling was that heavily overlapping objects which produce large
eigenvalues were annihilating. This na\"{\i}ve hope is seen to be
unrealized.  Cooling will also therefore alter the quark condensate,
as we see in figure~\ref{fig:4.xx.qc}.  (As usual this plot excludes
the exact zero modes which would give a finite-volume peaking of the
condensate at small quark masses.)  Whilst we should not pay too much
attention to the absolute normalisation of the quark condensate (given
the qualitative nature of the calculation), our observation that
cooling rapidly alters the quark condensate should be reliable.  This
creates an ambiguity that is particularly acute in the context of the
small-$\lambda$ divergence: depending on the amount of cooling, the
instanton ensemble produces a divergence in quenched QCD that ranges
from being very strong to being negligibly weak. The clear message is
that these instanton ensembles differ strongly in the long-distance
fermionic physics that they encode and that this is a problem that
needs to be resolved before one can be confident that one understands
the instanton content of the quenched QCD vacuum.

\begin{center}
\begin{figure}[p]
\leavevmode
\epsfxsize=100mm
\epsfbox{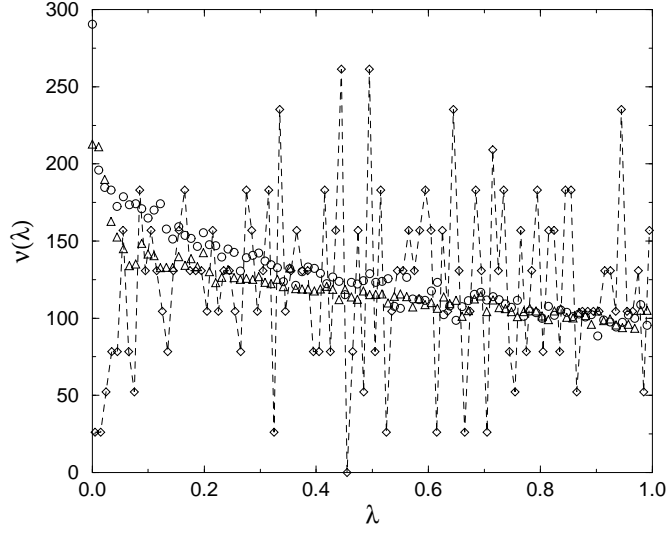}
\caption{Extending the calculations from actual lattice
data. ($\diamond$) $\beta = 6.4\ 32^{3}64$ data. $(\circ),\
(\triangle)$ from synthetic ensembles with approximately four times
the volume, and half the volume respectively.}
\label{fig:h_latxtend}
\end{figure}
\end{center}

It would be nice to have a study of the large volume limit at $\beta =
6.4$, similar to the one at $\beta= 6.0$. Unfortunately that would
require lattices much larger than $32^3 64$ and this is clearly
impractical. By contrast, a nice feature of using our model is that it
is easy to increase the volume and number of configurations and so
test whether one has reached the infinite volume limit (and to obtain
some idea of what a high statistics spectrum would look like).  We
show the results of such a calculation in
figure~\ref{fig:h_latxtend}. We compare the spectral density generated
from the instanton ensembles obtained after 80 cooling sweeps at
$\beta = 6.4$ to that from high statistics synthetic ensembles with
approximately half the volume and four times the volume
respectively. The packing fraction has been chosen to equal that of
the lattice ensemble. The lattice and model ensembles differ in that
the latter contain objects of a single size positioned at random and
with a total charge that is always zero, $Q=0$. We observe however
that the model spectra compare quite well with the lattice
spectrum. One difference is that the lattice spectrum lacks a forward
peak but this is in part due to the fact that the model ensemble
always has $Q=0$ while the lattice configurations do not. We note from
the figure that the two volumes produce essentially identical (model)
spectra.  Thus the $V\to\infty$ limit appears to be under control.

Finally we address the question of scaling. In \cite{DSMT} it was
shown that if we vary the number of cooling sweeps with $\beta$
appropriately, then many properties of the instanton ensemble become
independent of $\beta$ once they are expressed in physical units. Is
this also true of the more subtle features that are embodied in
physical observables such as the chiral condensate?  To investigate
this we plot in figure~\ref{fig:x.xx.scl} the spectral densities
obtained after 23, 46 and 80 cooling sweeps on the $16^3 48$, $24^3
48$ and $32^3 64$ lattices at $\beta\ = 6.0,\ 6.2,\ 6.4$
respectively. These lattices have nearly equal volumes in physical
units and the variation with $\beta$ of the number of cooling sweeps
is as prescribed in \cite{DSMT}.  As we see the corresponding spectral
densities are very similar showing that the important fermionic
physical observables do indeed scale.

\section{CONCLUSIONS}

In this paper we have investigated the implications of topology for
the chiral condensate in quenched QCD. For the topological structure
of the quenched QCD vacuum we have used the results of \cite{DSMT},
where cooled lattice fields are decomposed into ensembles of
instantons, and this is done for a variety of lattice spacings,
space-time volumes and cooling sweeps.  In addition, the lattice
analysis has been supplemented by studying ensembles of instantons
generated by a model that places them at random on the hypertorus. To
calculate the spectral density of the Dirac operator we have radically
simplified the calculation keeping only the most essential symmetries
and restricting the space of states to that spanned by the would-be
zero modes of the topological charges. This means that while we may be
able to address qualitative features, such as the variation of the
spectral density with volume or with cooling, we would not expect more
quantitative aspects, such as the actual value of the chiral
condensate, to be accessible to our approach. (Although, as we have
seen, it does not do too badly even for this.) In addition to the
general features of the chiral condensate, a particular question we
wished to address was whether such fermionic observables are
insensitive to the cooling or not.

One qualitative feature that is common to all the instanton ensembles
that we have investigated, is that they lead to spontaneous chiral
symmetry breaking.  A second, and striking, qualitative feature is
that the spectral density diverges as $\lambda \to 0$
\cite{ndmt,dvgce-other,cpt-log,cpt-power}.  The divergence follows an
approximate power law, $\propto \lambda^{-d}$, where $d$ decreases as
the density of the instantons increases. Moreover we have seen that it
is possible to have a stronger divergence for denser gases if one has
a sufficient range of instanton sizes, of the kind that one finds in
the lattice instanton ensembles.  We have also tried to fit the
divergence with a logarithmic form, since this is what one expects in
leading-order quenched chiral perturbation theory \cite{cpt-log}.  (It
has also been seen by unfolding the microscopic spectral density
obtained via Random Matrix Theory \cite{dvgce-other}.)  However such
logarithmic fits are usually unacceptable, and where they are not it
is a trivial consequence of the power exponent $d$ being small, as in
equation~(\ref{eq:log-power}).  It is interesting to note that if one
attempts to sum the leading-logs of quenched chiral perturbation
theory, one can obtain \cite{cpt-power} a power divergence. The
exponent of this divergence is $d=\delta/(1+\delta)$ where the
parameter $\delta$ is simply related to the elementary pseudoscalar
flavour singlet annihilation diagram (whose iteration provides an
estimate of the mass of the $\eta^{\prime}$ in full QCD). The strength
of this diagram is related, in turn, to the topological structure of
the quenched vacuum \cite{q-eta}, and so this suggests an approach to
constructing a detailed link between our approach and that of chiral
perturbation theory. It is amusing to note that the most recent
quenched QCD estimates \cite{delta} of $\delta$, obtained from chiral
extrapolations where this parameter multiplies the quenched chiral log
term, suggest a value $\delta \sim 0.1$ which is consistent with the
kind of weak divergence we typically observe on the cooled instanton
ensembles (see Table~\ref{tab:cdata}). We intend to explore this link
elsewhere \cite{usprep}.

We have furthermore found evidence, from a comparison of the Dirac
spectral densities, for some of the claims in \cite{DSMT}: in
particular for the screening of topological charges in the quenched
QCD vacuum, for the smallness of finite volume corrections and for the
claim that if the number of cooling sweeps is varied with $\beta$ so
that the number of topological charges per unit physical volume is
constant, then the physical observables show scaling.

However we have also found that fermionic physical observables, such
as the chiral condensate, vary strongly with the number of cooling
sweeps.  This contradicts the expectation that a moderate amount of
cooling should only eliminate short distance fluctuations and so
should not alter the physically important small-$\lambda$ end of the
spectral density. Whether this is a problem with cooling {\it per se}
or whether, as one would expect, it indicates the increasing
unreliability of the instanton ``pattern recognition'' algorithms of
\cite{DSMT} as one decreases the number of cooling sweeps, is a
question we are not able to address. The resulting uncertainty is
particularly important for the significance of the small-$\lambda$
divergence. We have seen that this divergence ranges from being strong
to being negligible depending on which of the lattice instanton
ensembles is used. It is strong for the larger number of cooling
sweeps, which is where the instanton pattern recognition should be
more reliable. On the other hand, a recent analysis \cite{arfs}
suggests that it is the instanton ensembles obtained with less
cooling, where the low-$\lambda$ peaking is negligible, that are the
more physical.  So although we do find that instantons generically
produce a divergence in the chiral condensate of quenched QCD, it is
not clear whether it is strong enough to have any impact on the
predictions for physical quark masses.  One lesson is unambiguous:
there is more that needs to be done before one can claim to have
completely understood the true instanton structure of the gauge theory
vacuum.

\section*{Acknowledgements}

The lattice instanton ensembles used in this work were produced by
Douglas Smith and one of the present authors (MT). They were derived
from gauge field configurations produced by UKQCD. We are grateful for
this material. MT also wishes to acknowledge his debt to Nigel Dowrick
without whose essential contributions, in an earlier collaboration,
the present study would not have been possible.  We are also grateful
to various colleagues (including the referee) whose response to the
first draft of this paper has helped to improve it. US wishes to thank
PPARC for financial support (research studentship number
96314624). The computations herein were performed on our Departmental
workstations.  We are grateful to PPARC for support under PPARC grant
GR/K55752.

\begin{center}
\begin{table}[htb]
\begin{tabular}{|c|c|c|c|c|c|c|c|c|}
$\beta$ & $L^{3}T$ & a (fm) & Cools & $N_{c}$ &
$\overline{N}_{I}/V$ ($fm^{-4}$) &
$\overline{N}_{I}\overline{V}_{I}/V$ & d & $\chi^{2}/dof$\\ \hline
6.0 & $16^{3}48$ & 0.098 & 23 & 100 & 9.1 & 4.2 & - & -\\
6.0 & $16^{3}48$ & 0.098 & 46 & 100 & 3.2 & 1.9 & - & -\\
6.0 & $32^{3}64$ & 0.098  & 46 & 50 & 3.5 & 2.84 & 0.251 $\pm$ 0.069 & 0.16\\
6.0 ({\em RP}) & $32^{3}64$ & 0.098 & 46 & 50 & 3.5 & 2.84 & 0.585 $\pm$
0.047 & 0.22\\
6.0 ({\em SS}) & $32^{3}64$ & 0.098 & 46 & 50 & 3.5 & 1.85 & 0.186
$\pm$ 0.060 & 0.24\\
6.0 ($\rho > 0.12$) & $32^{3}64$ & 0.098 & 46 & 50 & 3.1 & 2.82 & - & -\\
6.2 & $24^{3}48$ & 0.072  & 46 & 100 & 8.9 & 4.9 & - & -\\
6.4 & $32^{3}64$ & 0.0545 & 30 & 20 & 56.6 & 12.4 & - & -\\
6.4 & $32^{3}64$ & 0.0545 & 50 & 20 & 21.7 & 8.5 & - & -\\
6.4 & $32^{3}64$ & 0.0545 & 80 & 20 & 9.2 & 5.3 & - & -\\
\end{tabular}
\vspace{0.2in}
\caption{Some information about the data analysed in this paper. $V$
is the total spacetime volume, $\overline{V}_{I}$ is the average
volume of an instanton in the ensemble, $\overline{N}_{I}$ is the
average number of topological charges per configuration and $N_{c}$ is
the number of configurations in the ensemble. {\em RP} stands for
random positioning of objects whilst {\em SS} stands for all the
objects being set to the same size. A degree of divergence
$d$ from a fit of the data to $\nu(\lambda) = a + b/\lambda^{d}$ is
given wherever a fit with reasonable statistical and systematic 
errors is possible. $\chi^{2}/dof$ is the $\chi^{2}$ per degree of
freedom: the standard measure of goodness of fit.}
\label{tab:cdata}
\end{table}
\end{center}

\begin{center}
\begin{table}[htb]
\begin{tabular}{|c|c|c|c|c|c|}
$f$ & $N_{I} + N_{\overline{I}}$ & $N_{c}$ & d & $\chi_{p}^{2}/dof$ &
$\chi_{l}^{2}/dof$\\ \hline
0.2 & 13 + 13 & 650000 & 0.656 $\pm$ 0.003 & 0.24 & 60\\
0.5 & 32 + 32 & 128000 & 0.695 $\pm$ 0.002 & 0.19 & 51\\
1.0 & 63 + 63 & 126000 & 0.595 $\pm$ 0.002 & 0.14 & 26\\
1.75 & 111 + 111 & 111000 & 0.309 $\pm$ 0.002 & 0.19 & 5\\
2.5 & 158 + 158 & 63200 & 0.075 $\pm$ 0.013 & 0.17 & 0.26\\
5.3 & 336 + 336 & 6720 &  0.013 $\pm$ 0.039 & 0.24 & 0.23\\
10.0 & 633 + 633 & 1266 & 0.009 $\pm$ 0.081 & 0.26 & 0.26\\
\end{tabular}
\vspace{0.2in}
\caption{Some information about the synthetic ensembles analysed in
this paper. $f$ is the packing fraction in each configuration, $N_{I}
+ N_{\overline{I}}$ are the number of instantons and anti-instantons
in each configuration respectively, $N_{c}$ is the number of
configurations in the ensemble, $d$ is again the degree of divergence,
and $\chi_{p}^{2}/dof$ is the standard chi-squared measure of goodness
of fit for the power law fit, $\chi_{l}^{2}/dof$ is the chi-square for
the log fit $\nu(\lambda) = a + b\ln(\lambda)$.}
\label{tab:sdata}
\end{table}
\end{center}

\end{document}